\documentclass[twocolumn,secnumarabic,amssymb, nobibnotes, aps, prd]{revtex4-1}

\usepackage{graphicx}
\usepackage{wrapft}
\usepackage{amsmath}
\usepackage{bm}
\usepackage{mathrsfs}
\usepackage{color}

\begin{document}

\title{Half-quantized Non-Abelian Vortices in Neutron 
$^3P_2$ Superfluids\\ inside Magnetars}

\author{Kota Masuda$^{1,2}$ and Muneto Nitta$^{3}$}
\email{masuda(at)nt.phys.s.u-tokyo.ac.jp, 
nitta(at)phys-h.keio.ac.jp}
\affiliation{$^{1}$Department of Physics, The University of Tokyo, Tokyo 113-0033, Japan\\
$^{2}$Theoretical Research Division, Nishina Center, RIKEN, Wako 351-0198, Japan\\
$^{3}$Department of Physics at Hiyoshi, 
and Research and Education Center for Natural Sciences,\\ Keio University,
Hiyoshi 4-1-1, Yokohama, Kanagawa 223-8521, Japan}
\date{\today}

\begin{abstract}
We point out that half-quantized non-Abelian vortices
exist as the minimum energy states in rotating neutron $^3P_2$ superfluids in 
the inner cores of magnetars with 
magnetic field greater than $3 \times 10^{15}$ Gauss, 
while they do not in ordinary neutron stars with smaller magnetic fields.
One integer vortex is split into two half-quantized vortices. 
The number of vortices is about $10^{19}$ 
and they are separated at about $\mu$m in a vortex lattice
for typical parameters, while 
the vortex core size is about 10-100 fm. 
They are non-Abelian vortices characterized 
by non-Abelian first homotopy group,
and consequently 
when two vortices corresponding to 
non-commutative elements collide, 
a rung vortex must be created between them,
implying the formation of an entangled vortex network 
inside the cores of magnetars.
We find the spontaneous magnetization in the vortex core 
showing anti-ferromagnetism 
whose typical magnitude is about $10^{8-9}$ Gauss 
that is ten times larger than that of integer vortices, 
when external magnetic fields are present 
along the vortex line.

\end{abstract}

\maketitle

\section{Introduction}

Neutron stars (NSs) or pulsars are 
rapidly rotating,  massive and compact stars 
accompanied by a strong magnetic field,
and so they provide a subject extensively studied 
in broad area of physics from 
astrophysics to nuclear and high energy physics 
and condensed matter physics.
Most abundant middle-aged pulsars show 
magnetic field $B \sim 10^{11-13}$ Gauss 
on their surfaces. 
In recent years, 
young NSs with much stronger magnetic field 
$B \sim 10^{13-15}$ Gauss on their surfaces   
have been observed 
by the observations of
some soft gamma ray repeaters and anomalous X-ray pulsars.
This class of NSs is called as magnetars \cite{Paczynski:1992zz,Thompson95,0004-637X-473-1-322,Melatos:1999ji}.
There have been various theoretical attempts to 
attribute the origin of the strong magnetic field of NSs 
to the intrinsic magnetization of the neutron star matter.
However, 
no enough explanation about their origin has been given. 
Here, we discuss consequences  
of such strong magnetic fields in magnetars
in the presence of neutron superfluids
in their cores. 

Neutrons  are believed to form Cooper pairs 
to constitute a superfluid 
inside the cores of neutron stars 
since the proposal \cite{Migdal:1959}, 
as in liquid helium 3, 
metallic superconductors, 
and ultracold fermion gases.
At  densities lower than the normal nuclear matter density, 
the conventional and isotropic singlet ($^1S_0$) pairs are formed
while the anisotropic triplet ($^3P_2$) attractive interaction becomes comparable to the $^1S_0$ pairs at the normal nuclear matter density, 
where a transition is expected to occur  
\cite{tamagaki:1970,PhysRevLett.24.775,takatsuka:1971,takatsuka:1972,
fujita:1972,Richardson:1972xn}.  
The widely accepted observational evidence 
for the existence of neutron superfluids 
is observed long relaxation time $\tau$ ($\sim$ weeks for Crab and $\sim$ years for Vela) after pulser glitches \cite{glitch-1,glitch-2,glitch-3}.
Other signals contain 
the sudden speed-up events of neutron stars 
(called pulsar glitches \cite{Reichley1971})
proposed to be a consequence of    
the unpinning dynamics of a large number of 
superfluid vortices pinned on the nuclei \cite{Anderson:1975zze},  
and the cooling process of a neutron star 
 \cite{Heinke:2010cr,Page:2010aw}.  
One of the most important consequences of 
the existence of a superfluid would be 
superfluid vortices.
If a superfluid is rotating inside a neutron star, 
it must be threaded by 
superfluid vortices along the rotation axis, 
as well established in 
helium superfluids and ultracold atomic gases. 
The number $N_v$ of vortices with the unit circulation 
created inside rotating neutron stars can be estimated to be
\begin{eqnarray}
N_v \sim 1.9\times 10^{19} \left(\frac{1{\rm ms}}{P}\right)\left(\frac{M^*}{900{\rm MeV}}\right)\left(\frac{R}{10{\rm km}}\right)^2 \label{eq:Nv}
\end{eqnarray} 
using the period  $P$ of the neutron star, 
 the effective neutron mass $M^*$ , 
and  the radius $R$ of the neutron superfluid. 
The vortices will constitute a lattice, 
and the mean distance between vortices is about 
$d \sim 1.7 \times 10^{-6} {\rm m}$ 
for the typical values for $P$, $M^*$ and $R$ in Eq.~(\ref{eq:Nv}),
which is much larger than the coherence length 
$\xi \sim 10-100$ fm of the neutron superfluid, 
or the core size of vortices. 
Therefore, a large number of thin vortices must exist.

Exotic vortices are predicted to exist in recent developments of 
ultracold atomic gases,
in particular in spinor Bose-Einstein condensates (BECs).  
Prime examples are 
fractionally quantized non-Abelian vortices 
\cite{Semenoff:2006vv,Kobayashi:2008pk,Kawaguchi:2012ii} 
in spin-2 BECs. 
The term ``non-Abelian'' indicates elements of the 
first homotopy group to be non-Abelian (non-commutative)
\cite{Poenaru:1976zz,Mermin:1979zz}, 
and consequently 
when two non-Abelian vortices, 
characterized by homotopy group elements that do not commute, 
collide, it is inevitable to form a rung vortex 
that bridges the two vortices \cite{Poenaru:1976zz,Mermin:1979zz,Kobayashi:2008pk},  
implying the formation of a tangled network of vortices 
and long lifetime of vortices after their formation 
at the phase transition \cite{Spergel:1996ai,McGraw:1997nx,Bucher:1998mh}. 
Such non-Abelian vortices 
may drastically change   statistical 
properties of superfluids in non-equilibrium, 
such as the Kolmogorov law of energy cascades of 
quantum turbulence. 

In this Letter, we point out that such non-Abelian vortices 
are present stably instead of conventional vortices 
in the neutron $^3P_2$ superfluids 
in the cores of neutron stars when they are 
accompanied by 
strong magnetic field larger than $3 \times 10^{15}$ Gauss 
corresponding to magnetars.  
In order to study the $^3P_2$ superfluids,  
the Ginzburg-Landau (GL) free energy, 
derived in Refs.~\cite{fujita:1972,Richardson:1972xn} 
in the weak coupling limit, 
is useful although it is
valid only near the critical temperature. 
For instance, 
the ground state was determined 
 to be in the nematic phase \cite{Sauls:1978lna} 
according to the classification by Mermin \cite{Mermin:1974zz}. 
In the absence of magnetic field,
it is continuously degenerated 
up to the forth order, 
as the nematic phase of spin-2 BECs \cite{Song:2007ca,Uchino:2010pf}. 
The ground state  
in the presence of the magnetic field
has been determined recently 
\cite{Masuda:2015jka} 
to be the uniaxial nematic phase 
with the magnetic field smaller than $10^{14}$ Gauss 
corresponding to ordinary neutron stars,   
the $D_2$ biaxial nematic 
with intermediate magnetic field, 
and the $D_4$ biaxial nematic phase 
with magnetic field stronger than $3 \times 10^{15}$ Gauss 
corresponding to magnetars. 
Integer vortex structures in $^3P_2$ superfluids were discussed 
in the GL equation in the absence 
\cite{Richardson:1972xn,Muzikar:1980as,Sauls:1982ie}
and  the presence  \cite{Masuda:2015jka} of magnetic fields. 
In particular, the spontaneous magnetization of a vortex core, 
pointed out in Ref.~\cite{Sauls:1982ie}, 
has been calculated explicitly \cite{Masuda:2015jka}. 
Here, we find half-quantized non-Abelian vortices 
in the $^3P_2$ superfluids 
in magnetars. 
In this case, the ground state is in the $D_4$ biaxial nematic 
phase. 
An integer vortex studied before 
is unstable to decay into two half-quantized non-Abelian vortices,
and therefore non-Abelian vortices are the most 
fundamental topological degrees of freedom. 
We classify vortices, construct the vortex solutions, 
and calculate the magnetization of the vortex core 
induced by the neutron anomalous magnetic moment,
with finding that it behaves as an anti-ferromagnet;
it is magnetized opposite to the direction 
of the applied magnetic field. 
The typical magnitude of the magnetic field 
inside the vortex core 
is about $10^{7-8}$ Gauss
that is ten times larger than that of 
integer vortices.

\section{Ginzburg Landau Free Energy for $^3P_2$ Superfluids and the Ground State}\label{sec:GL}

We first give the GL free energy 
and determine the ground states in 
the presence of strong magnetic fields.
The GL free energy for the $^3P_2$ superfluids 
was derived in 
Refs.~\cite{fujita:1972,Richardson:1972xn, sauls-thesis} 
in the weak coupling limit 
by considering only the excitations around the Fermi surface and
assuming the contact interaction.
The order parameter for $^3P_2$ superfluidity is given by 
$3 \times 3$ traceless symmetric tensor $A_{\mu i}$ 
defined by
\begin{eqnarray}
\Delta=\sum_{\mu i} i\sigma_{\mu}\sigma_y A_{\mu i} k_i
\end{eqnarray}
with the gap parameter $\Delta$.
Here, the Latin letter $\mu$ and 
the Roman letter $i$ 
stand for the spin index and spatial coordinates,  respectively. 
The continuous symmetry acting on the matrix $A$ is
\begin{eqnarray}
 A \to e^{i \theta} g A g^T, \quad e^{i \theta} \in U(1),
\quad g \in SO(3).
\end{eqnarray}
The free energy density $F$ 
can be written as
\begin{eqnarray}
F=\int d^3 \rho \ (f_{\rm grad} + f_{2+4}+f_6+f_H) \label{freeenergydensity}
\end{eqnarray}
where the gradient term $f_{\rm grad}$, 
the second, fourth $f_{2+4}$ and sixth order $f_6$  terms  \cite{sauls-thesis} and the magnetic interaction term  $f_H$ are given by 
\begin{eqnarray}
 f_{\rm grad} &=& K_1 \partial_iA_{\mu j}\partial_iA^{\dagger}_{\mu j} 
 + K_2(\partial_iA_{\mu i}\partial_jA^{\dagger}_{\mu j}+\partial_iA_{\mu j}\partial_jA^{\dagger}_{\mu i})  
\nonumber \\  
f_{2+4}&=&\alpha {\rm Tr}AA^{\dagger}
 +\beta[({\rm Tr}AA^{\dagger})^2-{\rm Tr}A^2A^{\dagger 2}],\nonumber \\
f_6 &=&
\gamma [-3({\rm Tr}AA^{\dagger})|{\rm Tr}AA|^2
+4({\rm Tr}AA^{\dagger})^3 \nonumber \\
&&+12({\rm Tr}AA^{\dagger}){\rm Tr}(AA^{\dagger})^2
+6({\rm Tr}AA^{\dagger}){\rm Tr}(A^2A^{\dagger 2}) \nonumber \\ 
&&
+8{\rm Tr}(AA^{\dagger})^3
+12{\rm Tr}[(AA^{\dagger})^2A^{\dagger}A] \nonumber \\
&&-12{\rm Tr}[AA^{\dagger}A^{\dagger}A^{\dagger}AA]
-12{\rm Tr}AA({\rm Tr}AA^{\dagger}AA)^{\ast}], \nonumber \\
f_H &=& g'_H H^2 {\rm Tr}(A A^{\dagger})+g_H H_{\mu}(AA^{\dagger})_{\mu \nu}H_{\nu},
\label{eq:fB}
\end{eqnarray}
respectively. 
The GL parameters are summarized 
in Appendix \ref{sec:GLpara}.

Following the classification of the ground states in the general case 
\cite{Mermin:1974zz}, the ground state of 
$^3P_2$ superfluids was found to be in the nematic phase 
\cite{Sauls:1978lna},
in which 
the tensor $A$ is in the form of 
$A \propto {\rm diag} (r,-(1+r),1)$ with a real parameter 
$r \in \mathbb{R}$ 
($-1 \leq r \leq -1/2$).
The ground state in the presence of 
external magnetic field $H$ 
was determined recently \cite{Masuda:2015jka}  
to be in the UN phase ($r=-1/2$)
or in $D_2$ BN phase ($-1<r<-1/2$)
for the weak magnetic field for ordinary neutron stars, 
and 
the $D_4$ BN phase ($r=-1$) 
for the strong magnetic field 
greater than $3 \times 10^{15}$ Gauss 
for magnetars.
Here, we consider such strong magnetic fields 
along the vortex line in the $z$ direction $(\bm{H} \parallel \bm{z})$
or the angular direction $(\bm{H} \parallel \bm{\theta})$.
In the ground state, the eigenvalues of the matrix $A$ 
are $1$, $-1$ and $0$, among which 
the zero eigenvalue is directed along
the magnetic field so that 
the energy contribution from the magnetic fields vanish.
Consequently, 
the order parameter  
\begin{eqnarray} \label{eq:magnetic_field} 
A_{\rm g.s.}
= \pm 
\sqrt{|\alpha|/2\beta}\,
{\rm diag.}\, (1,-1,0)
\label{eq:g.s.}
\end{eqnarray}
is diagonalized in the $(x,y,z)$ coordinate basis 
for $\bm{H} \parallel \bm{z}$ 
and in the $(\rho,z,\theta)$ basis 
for $\bm{H} \parallel \bm{\theta}$. 
For the latter case, non-zero components are 
$A_{\rho \rho} = - A_{zz} = \pm \sqrt{|\alpha|/2\beta}$,
and the order parameter in the original Cartesian 
$(x,z,y)$ coordinate basis can be obtained by 
$O (\theta)A_{{\rm g.s.}}O^T (\theta)$
with 
\begin{eqnarray}
O(\theta)=
\left(
    \begin{array}{ccc}
      {\rm{cos}}  \theta & 0 & -{\rm{sin}}  \theta \\
      0 & 1 & 0 \\
       {\rm{sin}}  \theta & 0 & {\rm{cos}}  \theta \\
    \end{array}
  \right).\label{eq:On}
\end{eqnarray}

The symmetry of this state is $D_4$ defined by 
\begin{eqnarray}
&&D_4 =\{ (1,{\bf 1}_3), (1,I_1), (1,I_2), (1,I_3), 
 (-1,R),  \nonumber\\ && \quad
(-1,I_1R), (-1,I_2R),(-1,I_3R) \}  
\subset U(1) \times SO(3), \nonumber \\
&&I_1 = \left(\begin{array}{ccc}
1 &    &\\
   &-1&\\
   &    &-1
\end{array}\right)
\quad 
I_2 = \left(\begin{array}{ccc}
-1&    &\\
   &  1&\\
   &    &-1
\end{array}\right),  \nonumber \\
&&
I_3= \left(\begin{array}{ccc}
-1&    &\\
   &-1&\\
   &    &1
\end{array}\right),
\quad 
R= \left(\begin{array}{ccc}
 0& -1&\\
 1&  0&\\
   &    &1
\end{array}\right) .\label{eq:D4}
\end{eqnarray}
Here, $I_{1,2,3}$ represent for $\pi$ rotations 
around the first, second and third axes, respectively,  
where the labels represent for
$(1,2,3) =(x,y,z)$ for 
$\bm{H} \parallel \bm{z}$ and 
$(1,2,3) =(\rho,z,\theta)$ for 
$\bm{H} \parallel \bm{\theta}$. 
The element $R$ represents for $\pi/2$ rotation 
around the the third ($z$ or $\theta$)-axis  
($R^2=I_3$) and is accompanied by 
a $\pi$ phase rotation of $U(1)$. 
The same $D_4$ BN phase appears in spin-2 BECs  
\cite{Song:2007ca,Uchino:2010pf,Kawaguchi:2012ii},
and so these systems share common features.
The order parameter space (OPS) of the $D_4$ BN phase is
\begin{eqnarray} 
G/H = {U(1) \times SO(3) \over  D_{4}} 
\simeq  {U(1) \times SU(2) \over  D_{4}^*} 
\label{eq:OP}
\end{eqnarray}
where $D_{4}^*$ is the universal covering group of $D_4$, given by
\begin{eqnarray} 
&& D_4^* = \{(1,\pm {\bf 1}_2), (1,\pm i \sigma_1),
 (1,\pm i \sigma_2), (1,\pm i \sigma_3), \nonumber\\
&&\quad\quad (-1,\pm C), (-1,\pm i\sigma_1 C), 
(-1,\pm  i \sigma_2 C),  
 (-1,\pm  i \sigma_3 C)
\},  \nonumber \\
&& \quad C = {1\over \sqrt 2} 
\left(\begin{array}{cc}
1+ i & 0 \\
 0 & 1-i
\end{array}
\right), \quad  (C^2= i \sigma_3)  \label{eq:D4*}
\end{eqnarray}
with the Pauli matrices $\sigma_a$. 
The nontrivial homotopy groups of the OPS in Eq.~(\ref{eq:OP})
up to the 4th are 
\begin{eqnarray} 
 \pi_1 = {\mathbb Z} \times_h D_4^*, \quad
 \pi_3 = {\mathbb Z}, \quad 
 \pi_4 = {\mathbb Z}_2.
\end{eqnarray}
We focus on  
the first homotopy group 
characterizing vortices, that contains non-Abelian group $D_4^*$
(see Ref.~\cite{Kobayashi:2011xb} for 
the definition of the product ``$\times_h$").
The first four elements constitute the quaternion group 
${\mathbb Q}$ for spin vortices in biaxial nematic liquid 
\cite{Volovik:1977zz,Mermin:1979zz,Preskill:1990bm}, 
while the rests correspond to half-quantized non-Abelian vortices.
We note that 
when two vortices characterized by the 
elements $a$ and $b$ collide, 
a vortex characterized by $[a,b]$ is created 
when it is a nonzero element and 
bridges these two vortices \cite{Mermin:1979zz}.

\section{Half-quantized Vortex} \label{sec:vortex}

Here we construct a half-quantized vortex 
corresponding to $(-1,R)$ in Eq.~(\ref{eq:D4}) 
and $(-1,\pm C)$ in Eq.~(\ref{eq:D4*}).
We consider the following Ansatz for the order parameter 
of a vortex state
\begin{eqnarray}
&& A
= \sqrt{\frac{|\alpha|}{6\beta}} e^{i \theta/2}  
R(\theta) 
  \left(
    \begin{array}{ccc}
      f & ige^{im\theta +i\delta} & 0 \\
      ige^{im\theta+ i\delta} & -f & 0 \\
      0 & 0 & 0
    \end{array}
  \right)  R^T(\theta) 
\nonumber \\
&& \quad \quad R(\theta)=
\left(
    \begin{array}{ccc}
      {\rm{cos}}\theta/4 & -{\rm{sin}}\theta/4 & 0 \\
       {\rm{sin}}\theta/4 & {\rm{cos}}\theta/4 & 0 \\
      0 & 0 & 1
    \end{array}
  \right) ,
\label{eq:ansatz}
\end{eqnarray}
in the $(x,y,z)$ basis 
for $\bm{H} \parallel \bm{z}$ 
and 
 in the $(\rho,z,\theta)$ basis 
for $\bm{H} \parallel \bm{\theta}$, 
with the position dependence in 
the cylindrical coordinates $(\rho,\theta,z)$.
For the latter, the order parameters in the Cartesian $(x,z,y)$ coordinates 
is 
$O(\theta) A O^T (\theta)$ with Eq.~(\ref{eq:On}).
Here 
$m$ 
 is an integer and 
$f(\rho), g(\rho)$  are profile functions 
with  
the boundary conditions 
\begin{eqnarray}
(f,g)  \to  (\sqrt{3},0) 
 \mbox{ as } 
\rho \to \infty,   \quad
(0,0)  
  \mbox{ as } \rho \to 0.   \label{eq:bc}
\end{eqnarray}
The configuration in Eq.~(\ref{eq:ansatz}) corresponds to the element $(-1,R)$ of $\pi_1$ in Eq.~(\ref{eq:D4*}) 
because of $R(\theta)=R$ and $e^{i\theta/2}=-1$ at $\theta =2\pi$.
The overall phase $e^{i \theta/2}$ represent half-quantization. 
Although $e^{i \theta/2}$ is not single-valued, 
$R$ gives the minus sign 
with 
keeping the order parameter itself 
single-valued at $\theta = 2 \pi$.
The Ansatz in Eq.~(\ref{eq:ansatz}) is the most general 
for axially symmetric configurations where
the other components must vanish to 
be compatible with the group action.

\begin{figure*}[htbp]
  \begin{center}
    \begin{tabular}{ccc}
      \begin{minipage}{0.33\hsize}
        \begin{center}
          \includegraphics[clip, width=6cm]{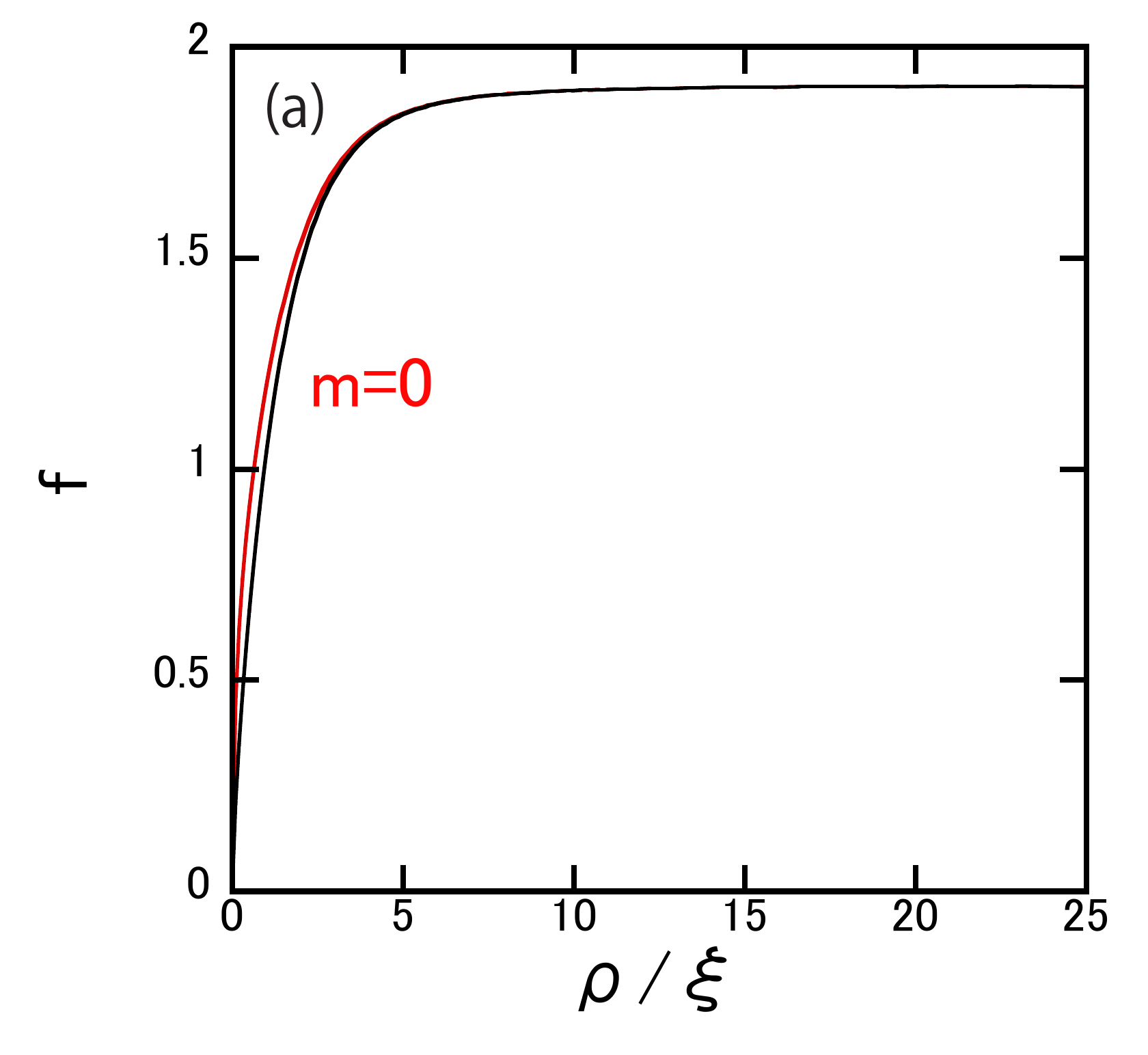}
                  \end{center}
      \end{minipage}
&      
      \begin{minipage}{0.33\hsize}
        \begin{center}
          \includegraphics[clip, width=6cm]{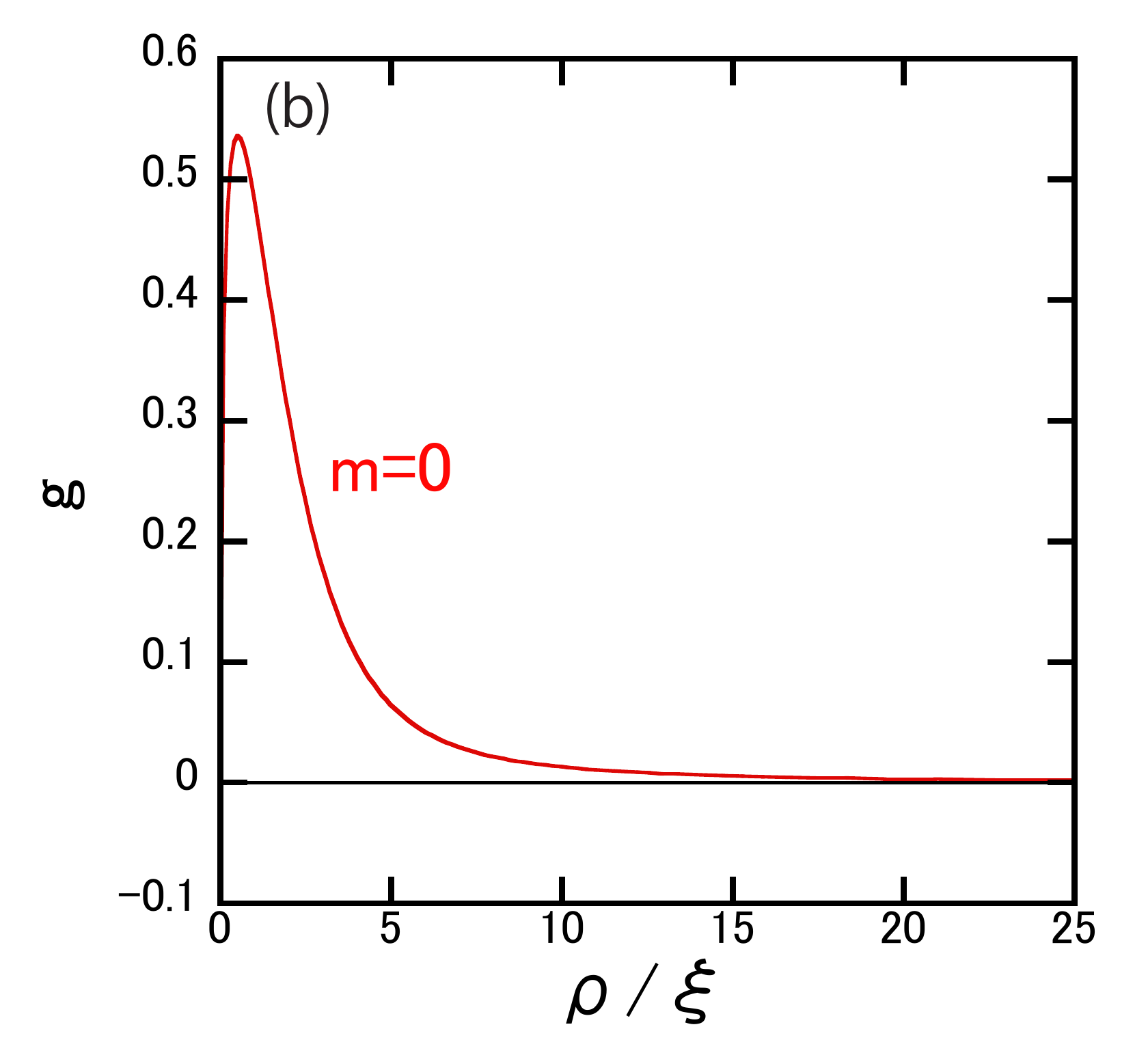}
        \end{center}
      \end{minipage}
&
      \begin{minipage}{0.33\hsize}
        \begin{center}
          \includegraphics[clip, width=6cm]{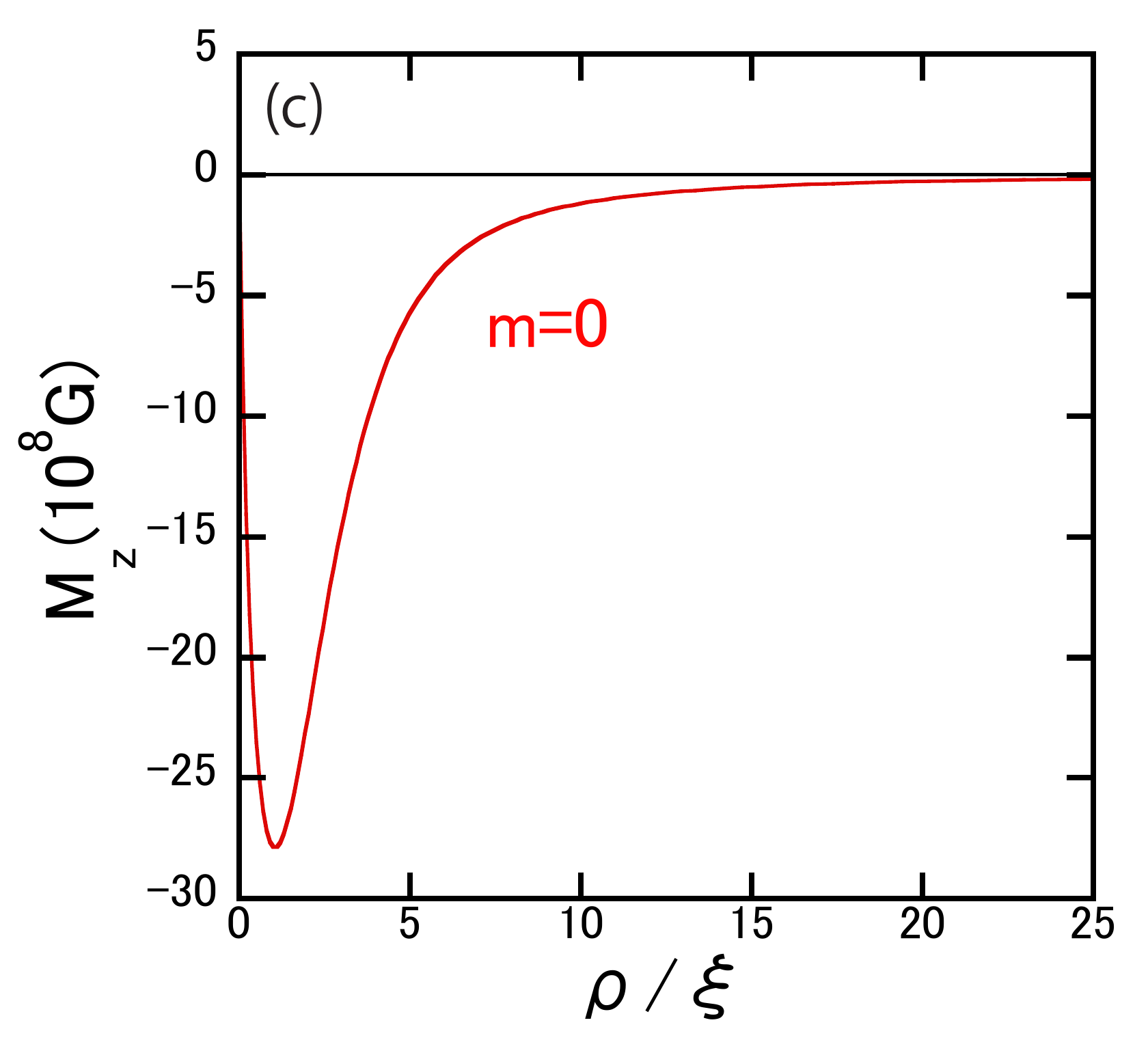}
        \end{center}
      \end{minipage}\\

      \begin{minipage}{0.33\hsize}
        \begin{center}
          \includegraphics[clip, width=6cm]{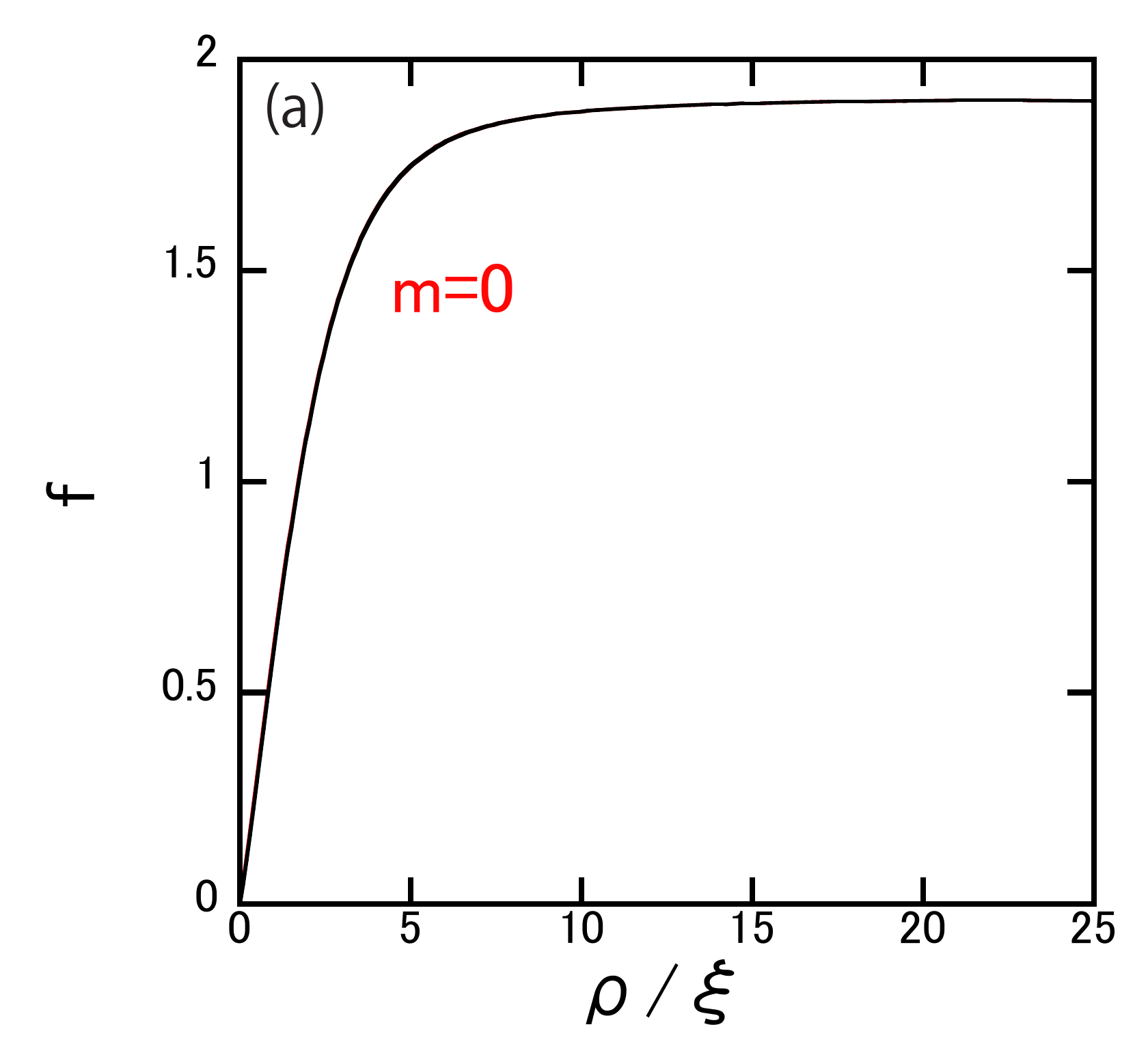}
                  \end{center}
      \end{minipage}
    &
      \begin{minipage}{0.33\hsize}
        \begin{center}
          \includegraphics[clip, width=6cm]{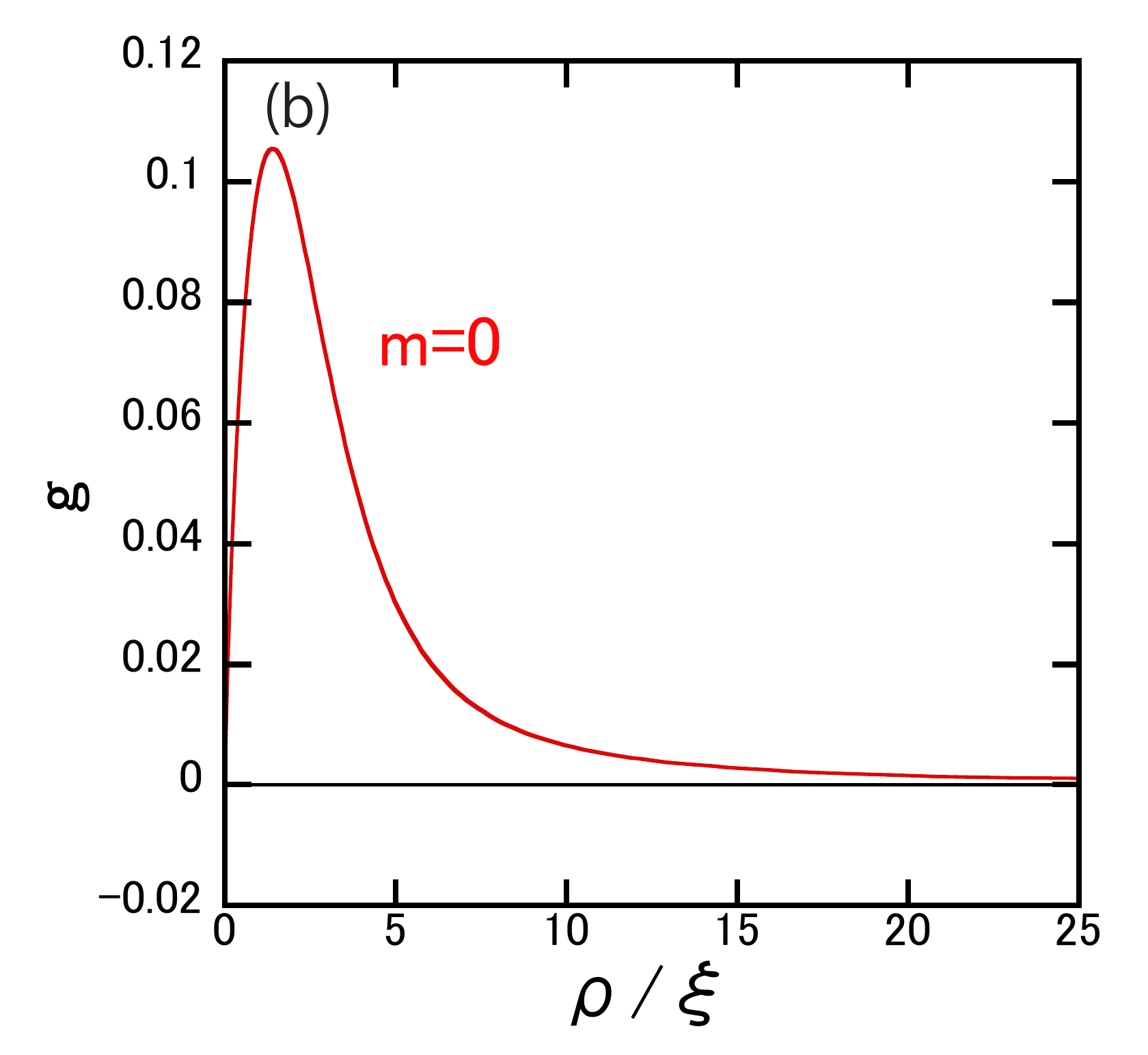}
        \end{center}
      \end{minipage}
   &
      \begin{minipage}{0.33\hsize}
        \begin{center}
          \includegraphics[clip, width=6cm]{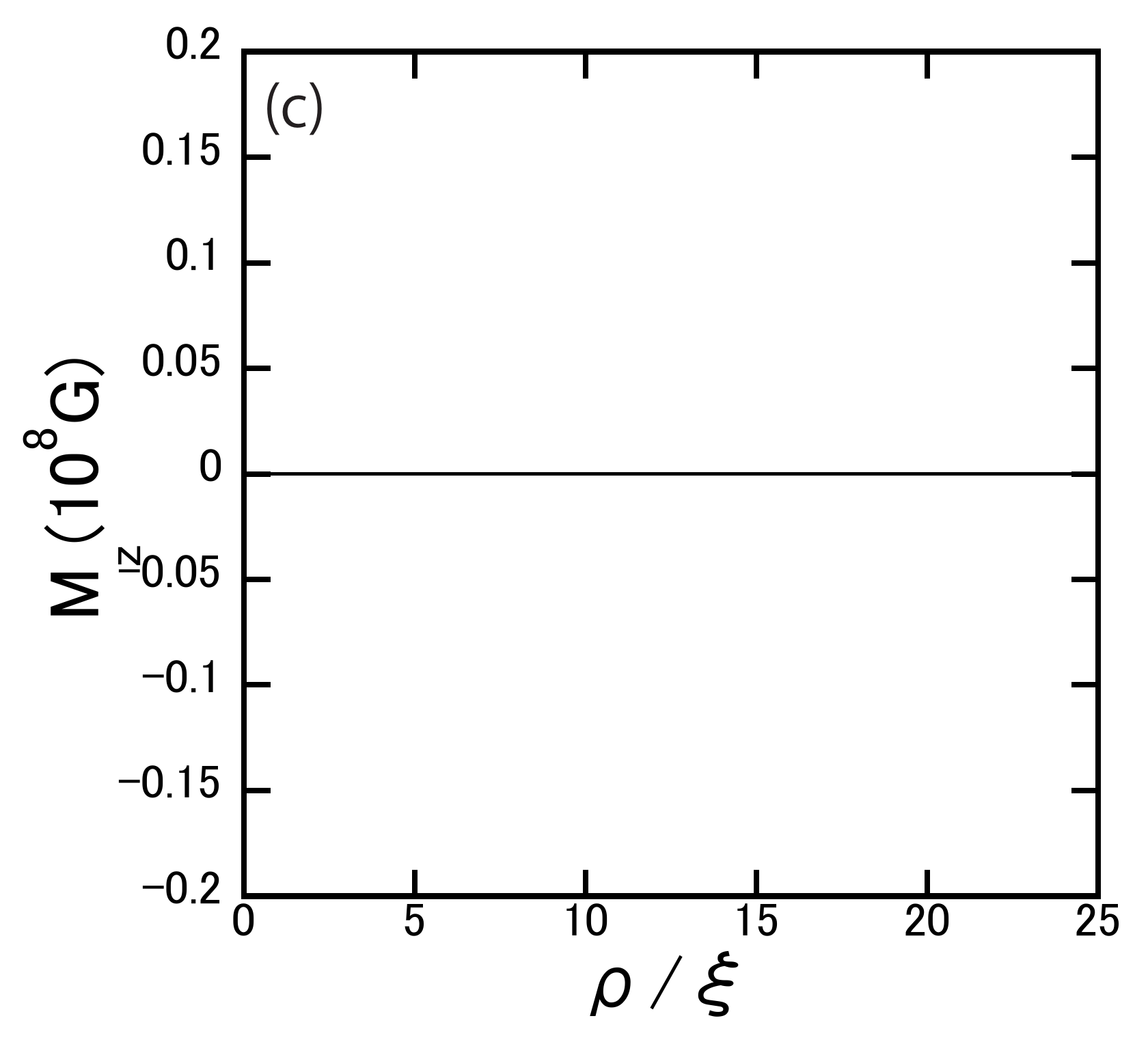}
        \end{center}
      \end{minipage} 
    \end{tabular}
    \caption{The profile functions and the magnetization. 
The upper and lower  panels are for 
$\bm{H} \parallel \bm{z}$ and 
$\bm{H} \parallel \bm{\theta}$, respectively.
The profile functions (a) $f$ and (b) $g$ as functions of the distance $\rho/\xi$ from the vortex center. 
The red and black curves correspond to 
$g \neq 0$ ($m=0$) and $g=0$ ($m \neq 0$) for which 
all $m$ result in the same solution. 
(c) The dependence of the magnetizations $M_z$ 
on the distance $\rho/\xi$ from the vortex core.
}
    \label{fig:profiles}
  \end{center}
\end{figure*}

We now solve vortex solutions numerically.
With the boundary conditions in Eq.~(\ref{eq:bc}), 
we solve the vortex profiles 
$f$ and $g$ as functions of the radial coordinate $\rho$ 
and plot in 
Fig.~\ref{fig:profiles} (a) and (b),
respectively. 
In numerical simulations,
we have changed $\rho$ ($0 \leq \rho < \infty$) as 
tanh$\rho$ ($0 \leq {\rm tanh}\rho < 1$), 
divided the domain of tanh$\rho$ into 100 parts
and solved the equations of motion 
in Appendix \ref{sec:eom} simultaneously in the Newton's method. 
We set 
the magnetic field (arbitrary) larger than $3\times 10^{15}$ Gauss,
for which the results do not depend on the value of the magnetic field.
In Fig.~\ref{fig:profiles} (a),
we plot the profiles $f$ with $m=0$ (the red curves) and $m \neq 0$ (the black curves) in the upper (for $\bm{H} \parallel \bm{z}$) 
and lower (for $\bm{H} \parallel \bm{\theta}$) panels. 
For both the cases, 
as shown by the equation of motions for the cylindrical basis in Appendix \ref{sec:eom},
Eq.~(\ref{eom-cyl-g}) is proportional to $g$ for $m \neq 0$.
Due to the boundary conditions $g=0$ at $\rho=0, \infty$,
only the trivial solution $g=0$ is allowed for $m \neq 0$
where all $m \neq 0$ give the identical solution.
For $\bm{H} \parallel \bm{\theta}$ in the lower panels,
the profiles $f$ with $m=0$ (the red curve) and $m \neq 0$ (the black curves) take different values although they are almost 
overlapped.
In summary, we have obtained two solutions 
$g \neq 0$ and $g=0$ for 
each case of  $\bm{H} \parallel \bm{z}$ 
and $\bm{H} \parallel \bm{\theta}$.
The solutions $g=0$ are metastable solutions.

Next, 
we calculate the spontaneous magnetization of $^3P_2$ vortex cores 
due to the neutron anomalous magnetic moment. 
The vortex magnetization $\bm{M}(\rho)$ in 
the half-quantized vortex core 
can be calculated as 
\begin{eqnarray}
\bm{M}&=&\frac{\gamma_n \hbar}{2} \bm{\hat{ \sigma}},
\end{eqnarray}
with the gyromagnetic ratio $\gamma_n$ of the neutrons 
and  
\begin{eqnarray}
\bm{ \hat{\sigma}}&=& T\sum_n \int \frac{d^3k}{(2\pi)^3}{\rm Tr}(\bm{\sigma}G(k,\omega_n))  \nonumber \\
&=&\frac{4}{9}N'(0)k_F^2\frac{|\alpha|}{6\beta}g(\rho)2f(\rho){\rm cos}m\theta \hat{\bm{z}}
\end{eqnarray}
where $G(k,\omega_n)$ is a thermal Green function and $\omega_{n}=(2n +1)\pi T$ is the Matsubara frequency 
and $N'(0) = {M^2 \over 2\pi^2 k_F}$ is the density of states differentiated 
by the energy $E=k^2/2M$, $N' = {M^2 \over 2\pi^2 k}$, 
evaluated at the Fermi surface $k=k_F$.  
We obtain the magnetization $\bm{M}$ as a function of $\rho$ 
and plot $M_z$
in Fig.~\ref{fig:profiles} (c). 
The red and black curves correspond to the cases for 
$g \neq 0$ ($m=0$) and $g=0$ ($m\neq 0$).
First, 
due to the axial symmetry around the $z$ axis,
the $(\rho,\theta)$ component of the tensor $A_{\mu i}$ in the cylindrical basis
or $(x,y)$ component of the tensor $A_{\mu i}$ in the Cartesian basis 
must have the nonzero value
to produce a net spontaneous magnetization along the $z$-axis.
In the case with $\bm{H} \parallel \bm{\theta}$,
the $(\rho,\theta)$ component of the tensor $A$ is zero as shown in
Eq.~(\ref{eq:ansatz}). Therefore, we obtain $M_z=0$ for all the cases.
On the other hand,
if we apply the magnetic field along the $z$-axis,
the $(x,y)$ component of the tensor $A$ 
is proportional to $g$.
As we already discussed,
$g$ has the nonzero value only for $m=0$.
This is the reason why only the case for $m=0$ has the spontaneous magnetization.
We have found that 
vortex core magnetization is anti-ferromagnetic,
that is, the magnetization is anti-parallel to 
the direction of the external magnetic field. 
The maximum value of $M_z$ is about $10^9$ Gauss,  
that is ten times larger than that with the integer vortex \cite{Masuda:2015jka}. 
This spontaneous magnetization is still negligible 
in observations, since 
the distance between vortices is much smaller than the coherence length
as we discussed in Ref.~\cite{Masuda:2015jka} 
for the case of an integer vortex. 

\begin{figure}[ht]
  \begin{center}
\vspace{-5mm}
          \includegraphics[clip, width=3cm]{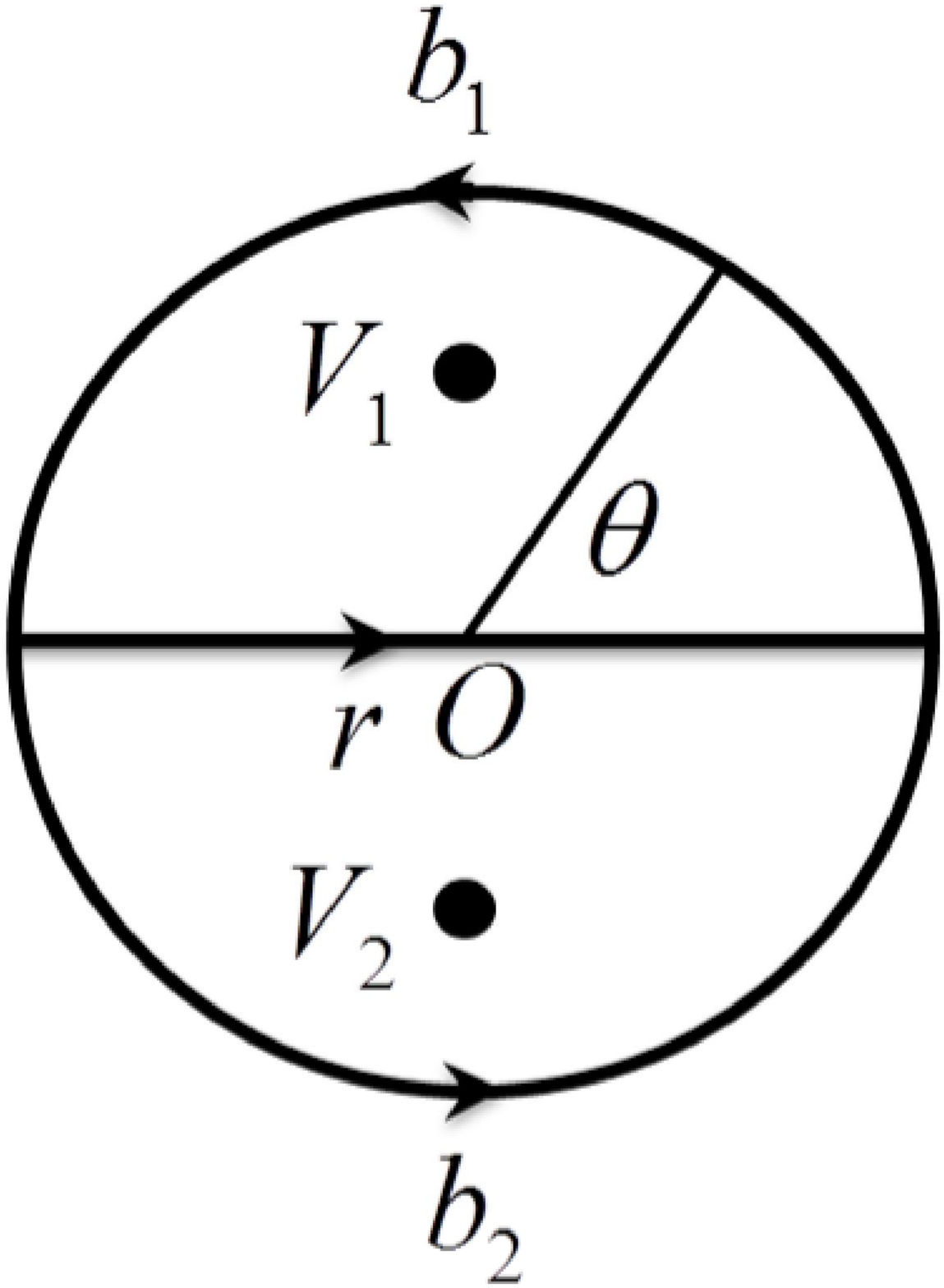}
\vspace{-10mm}
   \end{center}
\caption{Splitting an integer vortex into two half-quantized vortices.
\label{fig:splitting}}
\end{figure}

Finally, we discuss that half-quantized vortex states
give the minimum energy  in the $D_4$ BN phase under rotation 
rather than integer vortex states.
For simplicity we concentrate on the case of 
 $\bm{H} \parallel \bm{z}$.
Fig.~\ref{fig:splitting} shows how 
an integer vortex is split into two half-quantized vortices.
Let us consider an integer vortex located at the origin, 
given by 
$e^{i \theta}A_{\rm g.s.}$ 
with Eq.~(\ref{eq:g.s.})
at large distance $\rho \to \infty$
encircled by a path $b_1 + b_2$.
Along  each of the paths $b_1$ and $b_2$,  
the overall phase of the gap is rotated by $\pi$. 
Now let us consider a path $r$ along the $x$ coordinate 
that splits  $b_1 + b_2$. 
The integer vortex can be split into two half-quantized vortices separated 
in the $y$ direction 
at $V_1$ and $V_2$.
Along the path $r$, the tensor is given by 
$A = R(x) A_{\rm g.s.} R^T(x)$
where we assign an $SO(3)$ rotation 
$R(x)$ by $\pi/2$ given by 
\begin{eqnarray}
R(x)=
\left(
    \begin{array}{ccc}
      {\rm{cos}}\alpha(x) & - {\rm{sin}}\alpha(x) & 0 \\
       {\rm{sin}}\alpha(x) & {\rm{cos}}\alpha(x) & 0 \\
      0 & 0 & 1
    \end{array}
  \right) ,
\end{eqnarray}
with a real function  $\alpha(x)$ satisfying 
the boundary conditions 
$\alpha(x) \to 0$ at $x \to -\infty$ and 
$\alpha(x) \to \pi/2$ at $x \to +\infty$.
The closed paths $b_1 +r$ and $b_2 -r$ 
encircling $V_1$ and $V_2$, respectively, 
represent half-quantized vortices 
corresponding to $(-1,R)$ and $(-1,R I_3)=(-1,-R)$, 
respectively in Eq.~(\ref{eq:D4*}).
When one of them is well separated from the other, 
it gives the configuration in Eq.~(\ref{eq:ansatz}) 
where both the phase and $SO(3)$ rotation become 
monotonic to decrease the energy.   
This splitting is energetically favored because 
the tension of the vortex 
is proportional to the circulation [the $U(1)$ winding number] squared.
If the two half-quantized vortices are infinitely separated 
the energy is proportional to $(1/2)^2 + (1/2)^2 =1/2$ 
that is less than $1$ for one integer vortex.

\section{Summary and Discussion} \label{sec:summary}

In summary, half-quantized non-Abelian vortices 
exist as the minimum energy configurations 
in rotating neutron $^3P_2$ superfluids in 
the inner cores of magnetars,
in which the $D_4$ BN phase is realized.  
These vortices are superfluid vortices carrying 
half-quantized circulations and about $10^{19}$ vortices 
are created along the rotation axis 
for typical magnetars. 
When the magnetic field is present along the vortex line,
the spontaneous magnetization occurs in 
the vortex core, exhibiting anti-ferromagnetism 
of the order $10^{8-9}$ Gauss  
that is ten times larger than that of integer vortices. 
It does not occur for magnetic field encircling the vortex line.
These voritces belong to non-Abelian elements of 
the homotopy group $\pi_1$ and consequently 
a bridge must be created between them when they collide.  
Therefore the existence of an entangled network of vortices 
is predicted, 
implying long life time of vortices created at the phase transition.
When the magnetic field is weaker than $10^{15}$ Gauss 
for ordinary neutron stars,
the phase is in the $D_2$ or UN phase 
where half-quantized non-Abelian vortices do not exist. 
The existence and absence of 
non-Abelian vortices may characterize the distinct dynamics 
of magnetars and ordinary neutron stars. 
The nematic phase, 
including $D_4$ and $D_2$ BN and UN phases, also exists in spin-2 BECs 
of ultracold atomic gases, in which case these sub-phases are controllable experimentally, 
and so this opens a possibility to test certain aspects of physics of 
neutron stars in laboratory experiments.

\begin{acknowledgments}

We thank Takeshi Mizushima for helpful discussions.
KM thanks Mark Alford for the kind hospitality and discussions in Washington University in St.~Louis
where part of this work was carried out under the support of ALPS Program, 
University of Tokyo.
KM is supported by JSPS Research Fellowship for Young Scientists.
The work of M.~N.~is supported in part by a Grant-in-Aid for
Scientific Research on Innovative Areas ``Topological Materials
Science'' (KAKENHI Grant No.~15H05855) and ``Nuclear Matter in Neutron
Stars Investigated by Experiments and Astronomical Observations''
(KAKENHI Grant No.~15H00841) from the the Ministry of Education,
Culture, Sports, Science (MEXT) of Japan. The work of M.~N.~is also
supported in part by the Japan Society for the Promotion of Science
(JSPS) Grant-in-Aid for Scientific Research (KAKENHI Grant
No.~25400268) and by the MEXT-Supported Program for the Strategic
Research Foundation at Private Universities ``Topological Science''
(Grant No.~S1511006).

\end{acknowledgments}

\bibliography{./3p2-ref.bib}

\appendix

\section{GL parameters}\label{sec:GLpara}

\begin{table*}[t!]
\begingroup
\renewcommand{\arraystretch}{3}
 \begin{tabular}{c|c|c|c|c} 
    $\alpha$ & $K_1=K_2$ & $\beta$ & $\gamma$ & $g_H$  \\ \hline 
    $\displaystyle{\frac{N(0)}{3}\frac{T-T_c}{T}k_F^2}$ & 
    $\displaystyle{\frac{7\zeta (3)}{240M^2}\frac{N(0)}{(\pi T_c)^2}k_F^4}$ & 
    $\displaystyle{\frac{7\zeta (3)}{60}\frac{N(0)}{(\pi T_c)^2}k_F^4}$ &
    $\displaystyle{-\frac{31}{16}\frac{\zeta (5)}{840}\frac{N(0)}{(\pi T_c)^4}k_F^6}$ &    
    $\displaystyle{\frac{7\zeta (3)}{24}\frac{N(0)}{(\pi T_c)^2}\frac{(\gamma_n \hbar)^2}{2(1+F)^2}H^2 k_F^2}$  \\ 
  \end{tabular}
\endgroup
\caption{
The GL parameters in the weak coupling limit. 
}
\label{table-coeffi}
\end{table*}

In Table~\ref{table-coeffi},
we summarize the coefficients (the GL parameters) 
of the GL free energy in Eq.~(\ref{freeenergydensity}),
that are calculated in 
the weak coupling limit 
by considering only the excitations around the Fermi surface   \cite{fujita:1972,Richardson:1972xn, sauls-thesis}.
In this limit, $K_1$ and $K_2$ take the same value. 
In the derivation of $\alpha$, we took $g={3\pi^2 \over M k_F^3}$ 
in order for the $T$ dependence of $\alpha$ to
become that of the BCS theory.
$k_F$ is the Fermi momentum defined by $k_F = \hbar c (3\pi^2 \rho)^{1/3}$ 
with the neutron density $\rho$, 
$N(0) \equiv \frac{M k_F}{2\pi^2}$ is the density of states 
$N={M k \over 2\pi^2}$ 
on the Fermi surface at $k=k_F$,  
$T_c$ is the critical temperature for the $^3P_2$ superfluidity,
$\gamma_n$ is the gyromagnetic ratio of the neutrons
and 
$F$ is the Fermi liquid correction about the Pauli spin susceptibility.

The Riemann zeta function $\zeta (n)$ 
defined by $\zeta (n) = \sum_{k=1}^{\infty}\frac{1}{k^n}$ 
takes the values $\zeta (3) \sim 1.202$ 
and $\zeta (5) \sim 1.037$. 

In this Letter, 
we have taken $F=-0.75$,  $T_c = 0.2$MeV, $T = 0.8 T_c$ and $\rho  = 0.17$/fm$^3$ for numerical simulations.

\section{The free energy of a vortex}\label{sec:freeenergy} 

We calculate the free energy per the unit length of a vortex 
in Eq.~(\ref{eq:ansatz}):
\begin{eqnarray}
F&=&\int d^2 \rho \ \frac{|\alpha|}{6\beta}
\big( K_1 t_1+K_{2} t_2+\alpha t_3  
+\frac{|\alpha|}{6\beta}\beta t_4+\frac{\alpha^2}{36\beta^2} \gamma t_5 \big)
\nonumber \\
\end{eqnarray}
where $t_1$ and $t_2$ are 
the gradient terms and, 
$t_3$, $t_4$ and $t_5$ are
the second, fourth and sixth order terms, respectively  
in the GL free energy density in Eq.~(\ref{freeenergydensity}).
Here $t_{1,2}$ can be written as
\begin{eqnarray}
t^{(x,y,z)}_1
&=&
2\left(f'^2+g'^2
\frac{\frac{1}{2}f^2+\left(\frac{1}{4}+\left(m+\frac{1}{2}\right)^2\right)g^2}{\rho^2}
\right) \nonumber \\
&&-\frac{2}{\rho^2}
(m+1)fg{\rm cos}(m\theta +\delta), 
\end{eqnarray}
\begin{eqnarray}
t^{(x,y,z)}_2
&=&
2\left(f'^2+g'^2
\frac{\frac{1}{2}f^2+\left(\frac{1}{4}+\left(m+\frac{1}{2}\right)^2\right)g^2}{\rho^2}
\right) \nonumber \\
&&-\frac{2}{\rho^2}
(m+1)fg{\rm cos}(m\theta +\delta) 
\end{eqnarray} 
for the configurations diagonalized in the $xyz$ basis,
and 
\begin{eqnarray}
t^{(\rho,\theta,z)}_1&=&
2(f'^2+g'^2) 
+ \frac{1}{\rho^2}
\left(
3f^2
+\left(\frac{5}{2}+2\left(m+\frac{1}{2}\right)^2\right)g^2 \right. \nonumber \\
&& \left. -2(m+1)fg{\rm cos}(m\theta +\delta)
\right), \\
\label{gradient-1}
t^{(\rho,\theta,z)}_2&=&
2(f'^2+g'^2)
+\frac{2}{\rho}(ff'+gg')+\frac{2}{\rho^2}(f^2+g^2) 
\label{gradient-2}
\end{eqnarray}
for the configurations diagonalized in the cylindrical basis.
The rest terms 
$t_{3,4,5}$ can be written in the both basis as 
\begin{eqnarray}
t_3&=&2(f^2+g^2),\\ 
t_4&=&2f^4+2g^4
+8f^2g^2
+4f^2g^2{\rm cos}2(m\theta +\delta),\\
t_5&=&
48f^6
+48g^6  +(432+288{\rm cos}2(m\theta +\delta))f^4g^2  \nonumber \\
&&+(432+288{\rm cos}2(m\theta +\delta))f^2g^4.
\end{eqnarray}

We consider the case with $\delta=0$ 
consistent with the equation of motion, 
in which the imaginary part of non-diagonal elements is 
directly coupled to the real part of diagonal element. 
However, the effect of $\delta$ is yet to be clarified.

By differentiating the total free energy with respect to $f$ and $g$, 
we obtain the sets of the equation of motions for each basis,
as summarized in Appendix \ref{sec:eom}.

\section{Equation of motion} \label{sec:eom} 

\begin{widetext}

Here, we write down the equation of motion explicitly 
in the cylindrical basis ($n=1$) and  $xyz$-basis ($n=0$).
\begin{itemize}
\item  The equation of motions for cylindrical basis ($n=1$) 
are given as follows:
\end{itemize}

\begin{eqnarray}
&&8\frac{\partial^2 f}{\partial \rho^2}
+\frac{8}{\rho}\frac{\partial f}{\partial \rho} 
-\frac{1}{\rho^2}\left(10f-2\delta_{m,0}g \right) 
+4f 
-\frac{f}{6}\left(8f^2+(16+8\delta_{m,0})g^2\right) \nonumber \\
&&-\frac{|\alpha|}{36\beta^2}\gamma 
\Bigl(
288f^5 
+g^2(1728+1152\delta_{m,0})f^3 
+g^4(864+576\delta_{m,0})f
\Bigr)
=0,
\end{eqnarray}
\begin{eqnarray}
&&8\frac{\partial^2 g}{\partial \rho^2}
+\frac{8}{\rho}\frac{\partial g}{\partial \rho} 
-\frac{1}{\rho^2}
\left(
\left(
9+4\left(m+\frac{1}{2}\right)^2
\right)g
-2\delta_{m,0}f \right) 
+4g 
-\frac{g}{6}
\left(8g^2+(16+8\delta_{m,0})f^2\right) \nonumber \\
&&-\frac{|\alpha|}{36\beta^2}\gamma 
\Bigl(
288g^5 
+f^2(1728+1152\delta_{m,0})g^3 
+f^4(864+576\delta_{m,0})g
\Bigr)
=0.
\label{eom-cyl-g}
\end{eqnarray}

\normalsize{
\begin{itemize}
\item  The equation of motions for the $xyz$-basis ($n=0$) 
are given as follows:
\end{itemize}
}

\begin{eqnarray}
&&8\frac{\partial^2 f}{\partial \rho^2}
+\frac{8}{\rho}\frac{\partial f}{\partial \rho} 
-\frac{4}{\rho^2}\left(f-\delta_{m,0}g \right) 
+4f 
-\frac{f}{6}\left(8f^2+(16+8\delta_{m,0})g^2\right) \nonumber \\
&&-\frac{|\alpha|}{36\beta^2}\gamma 
\Bigl(
288f^5 
+g^2(1728+1152\delta_{m,0})f^3 
+g^4(864+576\delta_{m,0})f
\Bigr)
=0,
\end{eqnarray}
\begin{eqnarray}
&&8\frac{\partial^2 g}{\partial \rho^2}
+\frac{8}{\rho}\frac{\partial g}{\partial \rho} 
-\frac{4}{\rho^2}
\left(
\left(
\frac{1}{2}+2\left(m+\frac{1}{2}\right)^2
\right)g
-\delta_{m,0}f \right) 
+4g 
-\frac{g}{6}
\left(8g^2+(16+8\delta_{m,0})f^2\right) \nonumber \\
&&-\frac{|\alpha|}{36\beta^2}\gamma 
\Bigl(
288g^5 
+f^2(1728+1152\delta_{m,0})g^3 
+f^4(864+576\delta_{m,0})g
\Bigr)
=0.
\end{eqnarray}

\end{widetext}

\end{document}